\PassOptionsToPackage{table}{xcolor}
\documentclass[conference]{IEEEtran}
\usepackage[T1]{fontenc}
\usepackage[utf8]{inputenc}
\usepackage{mathptmx}   
\usepackage{svg}
\newcommand{\figfont}{\fontfamily{phv}\selectfont}   
\usepackage{graphicx}
\usepackage{booktabs}
\usepackage{amsmath}
\usepackage{siunitx}
\usepackage{diagbox}
\usepackage{tikz}
\usetikzlibrary{arrows.meta,positioning,fit,calc}
\usepackage{xcolor}
\usepackage[hidelinks]{hyperref}
\usepackage{hyperref}
\hypersetup{colorlinks=true, urlcolor=blue!50!black, linkcolor=black, citecolor=black}
\usepackage{cite}
\usepackage{enumitem}
\usepackage{makecell}
\usepackage{array}
\usepackage{pifont}
\usepackage{soul}
\newcommand{\cmark}{\ding{51}}\newcommand{\xmark}{\ding{55}}\newcommand{\pmark}{$\circ$}
\setstcolor{red}
\def\smallerspacecaption{\vspace{-2mm}}

\usepackage{multirow}

\usepackage{setspace}
\graphicspath{{../figures/}}

\DeclareTextFontCommand{\textbfit}{\bfseries\itshape}

\usepackage{setspace}

\definecolor{cin}{HTML}{2A78D6}\definecolor{cinf}{HTML}{E8F0FB}
\definecolor{cllm}{HTML}{4A3AA7}\definecolor{cllmf}{HTML}{EEEBFA}
\definecolor{csim}{HTML}{1BAF7A}\definecolor{csimf}{HTML}{E6F6EF}
\definecolor{cref}{HTML}{8A8984}\definecolor{creff}{HTML}{F2F2F0}
\definecolor{cfb}{HTML}{E8590C}
\definecolor{heat}{HTML}{2A78D6}\definecolor{venA}{HTML}{F1EEFB}\definecolor{venO}{HTML}{E9F6F0}
\definecolor{mOpus}{HTML}{4A3AA7}\definecolor{mSonnet}{HTML}{EB6834}\definecolor{mHaiku}{HTML}{C2338A}
\definecolor{mGsixone}{HTML}{1BAF7A}\definecolor{mGsix}{HTML}{2A78D6}\definecolor{mLuna}{HTML}{EDA100}
\definecolor{gradeG1}{HTML}{5B9BD5}\definecolor{gradeG2}{HTML}{2E6DB4}\definecolor{gradeG3}{HTML}{1B3F73}
\definecolor{tierT1}{HTML}{1BAF7A}\definecolor{tierT2}{HTML}{4F8FDB}\definecolor{tierT3}{HTML}{4A3AA7}\definecolor{tierT4}{HTML}{E8590C}\definecolor{tierT4f}{HTML}{8A8984}
\newcommand{\gradebox}[1]{{\setlength{\fboxsep}{1.2pt}\colorbox{grade#1}{\textcolor{white}{\bfseries\figfont\tiny #1}}}}
\newcommand{\tierbox}[1]{{\setlength{\fboxsep}{1.2pt}\colorbox{tier#1}{\textcolor{white}{\bfseries\figfont\tiny #1}}}}

\begin{document}

\title{\textsc{AnalogPortBench}: Benchmarking LLMs on\\Analog Circuit Porting}

\author{\IEEEauthorblockN{Saeid~Rajabi, Aly~Moussa, Vishal~Saxena, Chengmo~Yang, and Satwik~Patnaik}
\IEEEauthorblockA{Department of Electrical \& Computer Engineering\\
University of Delaware, Newark, DE, USA\\
\{srajabi, amoussa, vsaxena, chengmo, satwik\}@udel.edu}}

\maketitle

\begin{abstract}
Analog circuit design still relies heavily on expert knowledge, and porting proven designs to new process nodes is a recurring bottleneck. 
As a lower supply reduces available voltage headroom, a working topology may require redesign or fail to meet specifications. 
While large language models (LLMs) have begun to generate and optimize analog circuits, existing benchmarks 
focus on one process node and assume feasible specifications.
To bridge this gap, we introduce \textsc{AnalogPortBench}, a benchmark of 54 tasks for porting ten circuits, from current mirrors to telescopic amplifiers, from a 0.18\,\textmu m process to the open-source SKY130 PDK. Tasks range from a direct port through a reduced supply and harder specifications to infeasible ones, and every feasible task has a reference design verified at five process corners. Evaluating six frontier LLMs at two reasoning levels, with and without simulation feedback, we find that models port simple circuits but rarely re-design headroom-sensitive amplifiers, seldom recognize infeasible specifications, and gain more from simulation feedback than from more samples or reasoning. Code and data are available at \href{https://github.com/saeidrajabii/AnalogPortBench}{github.com/saeidrajabii/AnalogPortBench}.
\end{abstract}


\section{Introduction}
Analog circuits connect every electronic system to the physical world, with amplifiers, references, current mirrors, and bias networks sitting between sensors, converters, and digital logic. 
Unlike digital blocks, they are still custom-designed, with an expert balancing gain, bandwidth, stability, power, and voltage headroom against transistor nonidealities. 
A recurring part of this process is \emph{porting}, in which a block that meets its specifications in one process node is moved to the next, often together with a lower supply. 
This process is non-trivial because, for example, when the supply drops from 1.8\,V to 1.2\,V, a stack of five devices that fit comfortably may leave no room for saturation, and the designer must re-bias, fold the stack, or concede that a specification is out of reach.

Large language models (LLMs) have become capable engineering assistants. 
They write and debug code~\cite{chen2021codex,jimenez2023swebench}, generate and verify register-transfer-level hardware~\cite{thakur2022verigen,chang2023chipgpt,liu2023verilogeval,lu2023rtllm,thakur2023autochip,rajabi2025stellar,pinckney2024verilogeval2, gubbi2025prompting,allam2024rtlrepo,pinckney2025cvdp}, and have begun to design analog circuits by generating netlists with simulator feedback~\cite{lai2024analogcoder,lai2025analogcoderpro}, discovering topologies~\cite{gao2025analoggenie,zhang2024analogxpert}, building netlist datasets~\cite{bhandari2024masalachai,mehradfar2024aicircuit}, sizing devices~\cite{yu2026autosizer,liu2025eesizer}, and reading schematics~\cite{shi2025amsbench,akbari2025circuitsense}.
Beyond designing new circuits, one task of high importance to industry is \emph{porting}, which moves a working design to a smaller process node, often with a lower supply, while keeping its specifications~\cite{zhi2025analogtransfer}.
Existing analog benchmarks score LLM designs by simulation, but mostly ask models to design or size circuits \emph{within one process node}. Porting from one process node to another raises two challenges. 
First, reduced \emph{voltage headroom} can make the original topology not work, so the AI model may need to change it. 
Second, some specifications are unattainable,
so the AI model should declare them \emph{infeasible} instead of returning a failing circuit~\cite{zhong2025impossiblebench,kirichenko2025abstentionbench}. In software, the most related task is version-aware code migration~\cite{wu2024versicode}, where working code must be re-targeted to an environment that breaks the old solution, and correctness is checked by execution. Analog porting adds a twist that software lacks, since \emph{physics decides whether a valid port exists at all}.

Before LLMs can be trusted to port analog designs, their ability to do so must be measured systematically.
To address this need, this work presents \textsc{AnalogPortBench}, a simulation-based benchmark for analog porting. In each task, an LLM ports a working 0.18\,\textmu m design to the open-source SKY130 process at a supply of 1.8\,V or 1.2\,V. 
The benchmark contains multiple task tiers that separate the effect of the process node from that of the supply voltage. It also includes tasks with infeasible specifications to test whether an LLM can recognize that no valid design exists.

Overall, this work makes the following contributions:
\begin{itemize}[leftmargin=*,itemsep=1pt,topsep=2pt]
\item \textsc{AnalogPortBench}, a porting benchmark of 54 tasks over ten circuits, organized in three circuit difficulty grades and four task tiers, with reference designs verified at five process corners and four infeasible tasks.
\item A systematic, open-source evaluation flow based on SKY130~\cite{sky130pdk} and ngspice~\cite{ngspice} that checks each generated design step by step, from netlist validity to every specification limit.
\item A study of six representative LLMs at two reasoning levels, with and without simulation feedback, that answers five research questions. \textbf{RQ1:}~Can LLMs port a working design? \textbf{RQ2:}~What breaks when headroom shrinks? \textbf{RQ3:}~Do they recognize infeasible specifications? \textbf{RQ4:}~Does more reasoning or simulation feedback help more? \textbf{RQ5:}~Where in the design flow do designs fail?
\end{itemize}

\section{Background and Related Work}
\label{sec:related}

\begin{figure*}[t]
\centering
\includegraphics[width=0.99\textwidth]{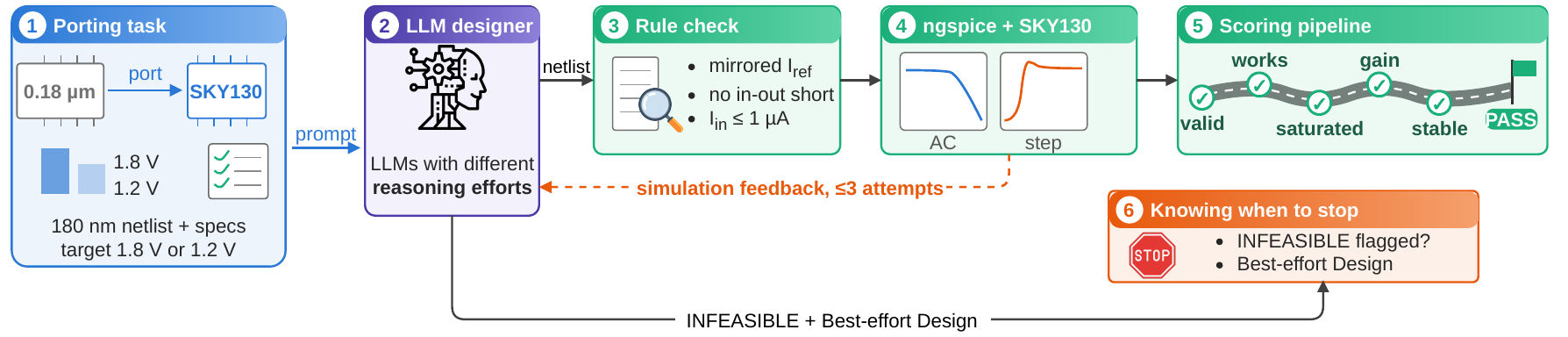}
\smallerspacecaption
\smallerspacecaption
\caption{Benchmark flow. The LLM receives the 0.18\,\textmu m netlist and the specifications; it checks the SKY130 netlist against the benchmark rules, simulates and scores it stage by stage, and scores INFEASIBLE answers on Tier T4 tasks. Simulation feedback allows up to three attempts.} \vspace{-6pt}
\label{fig:flow}
\end{figure*}

\noindent \textbf{LLMs for analog design.} Prior to LLMs, analog designs were optimized using Bayesian optimization~\cite{dong2023cktgnn} and reinforcement learning~\cite{wang2020gcn}. 
LLMs have enabled the generation of complete circuits, and recent benchmarks have evaluated them via circuit simulation.
Several LLM agents now design and size circuits with simulation feedback. AnalogCoder~\cite{lai2024analogcoder} generates circuits as Python code. 
EEsizer~\cite{liu2025eesizer} lets the LLM choose transistor dimensions directly, AutoSizer~\cite{yu2026autosizer} uses the LLM to guide a numerical optimizer, and AmpAgent~\cite{liu2024ampagent} combines an LLM with symbolic solvers and an external optimizer. 

\noindent \textbf{Porting.} Porting moves a working analog block to a new process node, often with a lower supply, while keeping its specifications. The designer re-sizes and re-biases each transistor, and changes the topology when the stack no longer fits the headroom.
Most LLM analog benchmarks evaluate designs within a single process (Table~\ref{tab:related}). AnalogCoder uses 24 circuits with simple transistor models, AMS-SizingBench uses 24 SKY130 circuits, and AmpAgent uses seven amplifiers in a 0.18\,\textmu m node. EEsizer is the only agent that moves a design across nodes, re-targeting a 20-transistor op-amp from 180\,nm to 90\,nm. However, all of them keep the circuit topology fixed, and none of them test whether a specification is feasible.

\begin{table}[t]
\caption{Comparison of AnalogPortBench with existing benchmarks.
Task tiers: one circuit at several difficulty levels;
Reasoning: reasoning levels compared.
\cmark{} yes, \pmark{} partially, \xmark{} no.}
\smallerspacecaption
\label{tab:related}
\centering\scriptsize
\setlength{\tabcolsep}{2.5pt}
\renewcommand{\arraystretch}{1.0}
\begin{tabular}{lcccc>{\columncolor{venA}}c@{}}
\toprule
\diagbox{\textbf{Feature}}{\textbf{Benchmark}}
 & \makecell{\textbf{Analog-}\\\textbf{Coder}\,\cite{lai2024analogcoder}}
 & \makecell{\textbf{Amp-}\\\textbf{Agent}\,\cite{liu2024ampagent}}
 & \makecell{\textbf{EEsizer}\\\cite{liu2025eesizer}}
 & \makecell{\textbf{AMS-}\\\textbf{Sizing}\,\cite{yu2026autosizer}}
 & \makecell{\textbf{This}\\\textbf{work}} \\
\midrule
Circuits/tasks      & 24/24  & 7/7    & 8/15   & 24/24  & \textbf{10/54} \\
Different difficulty levels   & \cmark & \xmark & \xmark & \cmark & \cmark \\
Different task tiers          & \xmark & \xmark & \pmark & \xmark & \cmark \\
Porting             & \xmark & \pmark & \pmark & \xmark & \cmark \\
Test for infeasibility & \xmark & \xmark & \xmark & \xmark & \cmark \\
Verified references provided      & \xmark & \xmark & \xmark & \xmark & \cmark \\
LLM reasoning studied & \xmark & \xmark & \xmark & \xmark & \cmark \\
\bottomrule
\end{tabular}
\end{table}
\section{\textsc{AnalogPortBench} Benchmark}

The proposed benchmark addresses the needs by testing whether an LLM can port a working design to a new process node and power supply, change topology when headroom requires it, and recognize when a specification cannot be met. Fig.~\ref{fig:flow} shows the 
flow. Each task gives an LLM a 0.18\,\textmu m design and a specification to meet in SKY130. The returned netlist is rule-checked and simulated, and the result is scored as a checklist or, for infeasible tasks, by whether the model declares infeasibility.

\textbf{Circuits and grades.} \textsc{AnalogPortBench} includes ten circuits from a 0.18\,\textmu m design set, 
grouped into three grades by
the voltage headroom and design effort required  (Table~\ref{tab:bench}):
\emph{G1 basic} (current mirrors, bias circuits, and a single-stage 5-transistor OTA), \emph{G2 medium} (two-stage Miller and folded-cascode OTAs) and \emph{G3 headroom-sensitive} (telescopic OTAs).

\begin{table*}[t]
\caption{Benchmark circuits grouped by difficulty grade, as well as the number of tasks per tier.}
\smallerspacecaption
\label{tab:bench}
\centering\footnotesize\setlength{\tabcolsep}{6pt}
\begin{tabular}{@{}clcl@{\hspace{6pt}}cccc@{}}
\toprule
\textbf{Grades} & \textbf{Circuit} & \textbf{FETs} & \textbf{Key specifications} & T1 & T2 & T3 & T4 \\
\midrule
\gradebox{G1} & Simple NMOS mirror & 2 & current error, $R_{out}$, $V_{min}$ & 1 & 1 & 2 & 1 \\
\gradebox{G1} & Cascode NMOS mirror & 4 & current error, $R_{out}$, $V_{min}$ & 1 & 1 & 2 & 1 \\
\gradebox{G1} & Cascode bias + P/N cascode source/sink & 10 & current error, matching vs.\ $V_{out}$& 1 & 1 & 2 & 1 \\
\gradebox{G1} & Wide-swing cascode bias + source/sink & 19 & current error, matching vs.\ $V_{out}$ & 1 & 1 & 3 & 1 \\
\gradebox{G1} & 5-transistor OTA with simple-mirror bias & 11 & gain, UGB, PM, power & 1 & 1 & 2 & -- \\
\midrule
\gradebox{G2} & Miller two-stage OTA, $C_c$--$R_z$ & 15 & gain, UGF, PM, GM, settling, power & 1 & 1 & 3 & -- \\
\gradebox{G2} & Miller two-stage, wide-swing bias & 24 & gain, UGF, PM, GM, settling, power & 1 & 1 & 3 & -- \\
\gradebox{G2} & Folded-cascode OTA, wide-swing bias & 31 & gain, UGB, PM, settling, swing, power & 1 & 1 & 3 & -- \\
\midrule
\gradebox{G3} & Telescopic cascode OTA, LV-cascode load & 17 & gain, UGF, PM, GM, settling, swing & 1 & 3 & 5 & -- \\
\gradebox{G3} & Telescopic OTA, replica-biased cascodes & 33 & gain, UGB, PM, settling, swing, power & 1 & 1 & 3 & -- \\
\midrule[0.8pt]
\rowcolor{heat!12} & \multicolumn{3}{l}{\textbf{All grades}\quad{\scriptsize 50 feasible + 4 infeasible = 54 tasks}} & \textbf{10} & \textbf{12} & \textbf{28} & \textbf{4} \\
\bottomrule
\end{tabular}
\\[2pt]
\parbox{\textwidth}{\scriptsize 
OTA: operational transconductance amplifier; LV: low-voltage; $C_c$--$R_z$: Miller capacitor with nulling resistor; current error: output-current error relative to the reference current; $R_{out}$: output resistance; $V_{min}$: compliance voltage; matching vs.\ $V_{out}$: source-sink current match over the output voltage range; UGB: open-loop unity-gain bandwidth; UGF: loop-gain unity-gain frequency; PM/GM: phase/gain margin; settling: step-response settling time; swing: output voltage range.}
\vspace{-9pt}
\end{table*}

\textbf{Task tiers.} The porting tasks of each circuit span up to four tiers (Table~\ref{tab:tiers}), with each tier adding extra difficulty, so that a failure can be traced to its cause. 
\textbf{T1} ports the circuit to SKY130 at the original 1.8\,V supply, testing whether an LLM can port a design to a new node when the topology still fits. 
Its limits come from the simulated source, with gain set to within 3\,dB of the source, bandwidth at least 0.8 times and power at most 1.2 times the source value, and a phase margin of at least 55$^\circ$. 
\textbf{T2} keeps the same specifications but lowers the supply to 1.2\,V and scales the voltage-range limits by 1.2/1.8. The drop from T1 to T2 thus isolates the effect of headroom. 
For the telescopic OTA, T2 instead has three alternative specification points that trade gain, speed, and swing. 
\textbf{T3} tests whether a model can re-design a circuit rather than only porting it. It sets 28 harder specification points at 1.2\,V, such as higher gain or lower compliance voltage, and each is met by a reference design to ensure feasibility.  
Finally, \textbf{T4} deliberately sets specifications just beyond a device's physics bound, so no circuit of the given type can meet them.
This limit holds because the task excludes output feedback. 
For example, a cascode current source and sink at 1.2\,V cannot keep their currents equal over more than 1.05\,V of output voltage (because each side needs some voltage near its rail), so asking for 1.06\,V is impossible.
T4 is designed to test if the LLM recognizes this and declares infeasibility. In total, the benchmark includes 54 tasks across tiers T1--T4 (Table~\ref{tab:tiers}), of which 50 are feasible tasks split across grades G1--G3, and four are infeasible tasks using G1 circuits (Table~\ref{tab:bench}).

\textbf{References and rules.} Each of the 50 feasible tasks has a reference design tuned by simulation. All of them meet every limit with every transistor in saturation at five process corners (TT, FF, SS, SF and FS), at 27\,$^\circ$C and the nominal voltage supply. Three benchmark rules prevent shortcuts in the netlist: the reference current must be copied through a mirror, no element may connect the input directly to the output, and the input may draw at most 1\,\textmu A.

\section{Experimental Setup}

\textbf{Models and settings.} We evaluate six LLMs, including Claude Opus, Sonnet and Haiku 5.5, all accessed through the command-line client with all tools disabled, as well as GPT-6-sol, GPT-6.1-sol and GPT-6-luna, all accessed through the OpenAI API. Each model runs at low and medium reasoning effort with default sampling, and each effort level answers each task three times independently, giving 1{,}944 zero-shot answers. Since effort settings are not comparable across vendors, we also record the output tokens of each answer, including reasoning tokens.

\textbf{Prompt.} Each prompt contains the 0.18\,\textmu m netlist, the specification table, the test conditions, the available SKY130 devices and the benchmark rules. The model sees the specifications and test conditions, but not the testbench files that measure them. The model may change transistor dimensions, biasing and topology, and must return one SKY130 netlist. If it decides that a task is infeasible, it must instead state INFEASIBLE, give the smallest relaxation that would make the task feasible, and still return a best-effort netlist. 

\textbf{Answer handling.} We extract the netlist from each answer, check it against the required ports and rules, and simulate it with the task's testbenches at the typical (TT) corner and 27\,$^\circ$C. We also check the operating point to confirm that every transistor is in saturation. Each answer is scored as \emph{pass} (every limit met), \emph{fail} (the circuit works but misses a limit), \emph{does not work} (no gain, or a $>$~50\% current error), or \emph{invalid} (no netlist, a rule violation, or a simulation error).

\textbf{Feedback mode.} At medium effort, every T1--T3 task whose first sample fails continues as a conversation with simulation feedback, with at most three attempts in total. After each attempt, the model receives the simulation results that include every metric with its limit, measured value, whether it was met, and the operating point of each transistor, but no hint about the cause of the failure.

\section{Results}

Table~\ref{tab:main} reports the zero-shot pass rate of each LLM per grade and tier, its INFEASIBLE flags on T4 and false alarms on feasible tasks, and the rate of tasks solved in three tries by resampling or by simulation feedback. 
As a baseline, the \emph{na\"ive port} replaces each 0.18\,\textmu m transistor with the SKY130 LVT device of the same width and length and changes nothing else. It passes only 2 of the 54 tasks, confirming that the benchmark cannot be solved by simply translating device names. The \emph{reference} line confirms feasibility of all T1-T3 tasks. 
The rest of this section answers the five research questions.

\begin{table}[t]
\caption{Task tiers, all target SKY130 1.8\,V low-threshold-voltage (LVT) devices, 54 tasks in total.}
\smallerspacecaption
\label{tab:tiers}
\centering\footnotesize\setlength{\tabcolsep}{3pt}
\renewcommand{\arraystretch}{0.92}
\setlength{\aboverulesep}{1pt}\setlength{\belowrulesep}{1pt}
\begin{tabular}{@{}cllc@{}}
\toprule
\textbf{Tier} & \textbf{$V_{DD}$, specification} & \textbf{Correct answer} & \textbf{Tasks} \\
\midrule
\tierbox{T1} & 1.8\,V, from simulated source & Porting meets all specification & 10 \\
\tierbox{T2} & 1.2\,V, voltage limits $\times$1.2/1.8 & Porting meets all specification & 12 \\
\tierbox{T3} & 1.2\,V, harder (reference exists) & Porting meets all specification & 28 \\
\tierbox{T4} & 1.2\,V, beyond a physical bound & INFEASIBLE + best effort & 4 \\
\bottomrule
\end{tabular}
\end{table}

\begin{table*}[t]
\caption{Zero-shot pass rate per tier and grade. Three samples per task, with task counts in parentheses. INFEASIBLE: answers declared infeasible for T4 tasks (correct, the higher the better) and feasible tasks (false alarms, the lower the better). Solved in 3 tries: rates of feasible tasks solved at medium effort by resampling (best of 3) or simulation feedback (up to 3 attempts). \\\colorbox{green!35}{Best} and \colorbox{green!15}{second-best} values in each column are highlighted.}
\smallerspacecaption
\label{tab:main}
\centering\footnotesize\setlength{\tabcolsep}{3.4pt}\renewcommand{\arraystretch}{0.8}
\begin{tabular}{@{}l@{\hspace{3pt}}c|ccc|ccc|ccc|rr|cc@{}}
\toprule
 & & \multicolumn{3}{c|}{\textbf{T1}} & \multicolumn{3}{c|}{\textbf{T2}} & \multicolumn{3}{c|}{\textbf{T3}} & \multicolumn{2}{c|}{\textbf{T4 / INFEASIBLE}} & \multicolumn{2}{c}{\textbf{Solved in 3 tries}} \\
\textbf{Model} & \makecell{\textbf{Reasoning}\\\textbf{effort}}
 & \multicolumn{1}{c}{G1\,{\scriptsize(5)}} & \multicolumn{1}{c}{G2\,{\scriptsize(3)}} & \multicolumn{1}{c|}{G3\,{\scriptsize(2)}}
 & \multicolumn{1}{c}{G1\,{\scriptsize(5)}} & \multicolumn{1}{c}{G2\,{\scriptsize(3)}} & \multicolumn{1}{c|}{G3\,{\scriptsize(4)}}
 & \multicolumn{1}{c}{G1\,{\scriptsize(11)}} & \multicolumn{1}{c}{G2\,{\scriptsize(9)}} & \multicolumn{1}{c|}{G3\,{\scriptsize(8)}}
 & \multicolumn{1}{c}{\makecell{Flagged\\{\scriptsize(4 tasks)} $\uparrow$}} & \multicolumn{1}{c|}{\makecell{False\\alarms $\downarrow$}}
 & \multicolumn{1}{c}{\makecell{Re-\\sampling}} & \multicolumn{1}{c}{\makecell{Simulation\\feedback}} \\
\midrule
\textit{Naive port} & -- & 20\% & 0 & 0 & 20\% & 0 & 0 & 0 & 0 & 0 & -- & -- & -- & -- \\
\midrule
\multicolumn{15}{@{}l}{\textit{OpenAI}} \\
\textcolor{mLuna}{\rule{5pt}{5pt}}\,GPT-6-luna & low & 40\% & 0 & 0 & 40\% & 0 & 0 & 0 & 0 & 0 & 6/12~ & 30/150~ & -- & -- \\
 & medium & 40\% & 0 & 0 & 47\% & 0 & 0 & 0 & 0 & 0 & 2/12~ & \cellcolor{green!15}{1}/150~ & 12\% & 12\% \\
\textcolor{mGsix}{\rule{5pt}{5pt}}\,GPT-6-sol & low & 53\% & \cellcolor{green!15}{11\%} & 0 & 47\% & 0 & 0 & 0 & 0 & 0 & 6/12~ & \cellcolor{green!15}{1}/150~ & -- & -- \\
 & medium & 73\% & \cellcolor{green!15}{11\%} & 0 & 53\% & 0 & 0 & 0 & 0 & 0 & 6/12~ & \cellcolor{green!35}{0/150}~ & 20\% & 34\% \\
\textcolor{mGsixone}{\rule{5pt}{5pt}}\,GPT-6.1-sol & low & 67\% & \cellcolor{green!15}{11\%} & 0 & 33\% & 0 & 0 & 6\% & 0 & 0 & 1/12~ & \cellcolor{green!35}{0/150}~ & -- & -- \\
 & medium & 73\% & 0 & \cellcolor{green!35}{17\%} & 53\% & 0 & \cellcolor{green!35}{8\%} & \cellcolor{green!35}{21\%} & 0 & 0 & 2/12~ & \cellcolor{green!35}{0/150}~ & \cellcolor{green!15}{30\%} & \cellcolor{green!35}{56\%} \\
\midrule
\multicolumn{15}{@{}l}{\textit{Anthropic}} \\
\textcolor{mHaiku}{\rule{5pt}{5pt}}\,Haiku 5.5 & low & 40\% & 0 & 0 & 27\% & 0 & 0 & 0 & 0 & 0 & 9/12~ & 20/150~ & -- & -- \\
 & medium & 27\% & 0 & 0 & 33\% & 0 & 0 & 0 & 0 & 0 & 9/12~ & 20/150~ & 8\% & 18\% \\
\textcolor{mSonnet}{\rule{5pt}{5pt}}\,Sonnet 5.5 & low & 60\% & 0 & 0 & 40\% & 0 & 0 & 0 & 0 & 0 & 6/12~ & 3/150~ & -- & -- \\
 & medium & 60\% & 0 & 0 & 40\% & 0 & 0 & 0 & 0 & 0 & 5/12~ & 4/150~ & 12\% & 26\% \\
\textcolor{mOpus}{\rule{5pt}{5pt}}\,Opus 5.5 & low & \cellcolor{green!15}{87\%} & 0 & 0 & \cellcolor{green!15}{60\%} & 0 & 0 & 0 & 0 & 0 & \cellcolor{green!35}{11}/12~ & 2/150~ & -- & -- \\
 & medium & \cellcolor{green!35}{93\%} & \cellcolor{green!35}{22\%} & 0 & \cellcolor{green!35}{73\%} & 0 & \cellcolor{green!35}{8\%} & \cellcolor{green!15}{15\%} & 0 & 0 & \cellcolor{green!15}{10}/12~ & \cellcolor{green!35}{0/150}~ & \cellcolor{green!35}{32\%} & \cellcolor{green!15}{52\%} \\
\midrule
\textit{Reference} & -- & 100\% & 100\% & 100\% & 100\% & 100\% & 100\% & 100\% & 100\% & 100\% & -- & -- & -- & -- \\
\bottomrule
\end{tabular}
\end{table*}

\textbf{RQ1. Can LLMs port a working design?} Partly, mainly for simple circuits. 
Although every LLM in Table~\ref{tab:main} beats the na\"ive device swap, most failing answers for T1 tasks are working circuits missing a limit rather than broken netlists. 
The best model Opus 5.5 passes 93\% of T1 answers on G1 circuits, but does not exceed 22\% on G2 or G3, even at the original supply.
Netlists returned by GPT models fail simulation more often than those by Claude models.

\textbf{RQ2. What breaks when the headroom shrinks?} Lowering the supply from 1.8\,V (T1) to 1.2\,V (T2) cuts the pass rate of the two strongest models 
from 53\% to 33\% for Opus~5.5 and from 40\% to 25\% for GPT-6.1-sol. The harder T3 tasks are rarely met by LLMs.
On G1 and G2 circuits, the stronger models still keep their transistors in saturation, but on the G3 headroom-sensitive telescopic cascode, only Opus~5.5 and GPT-6.1-sol do so, 
while the other models only re-size the telescopic stack. 
For this circuit, the 0.18\,\textmu m source stacks five transistors between the rails and swings only 0.54\,V, even at 1.8\,V. 
At 1.2\,V, Sonnet~5.5 resized it and reached about 0.23\,V, with three transistors in the triode region. 
Opus~5.5 noted that the stack cannot reach the limit with all transistors in saturation, replaced it with a two-stage Miller OTA, and met every limit. 
A correct porting here requires a topology change in addition to transistor re-sizing.

\begin{figure*}[t]
\centering
\includegraphics[width=\textwidth]{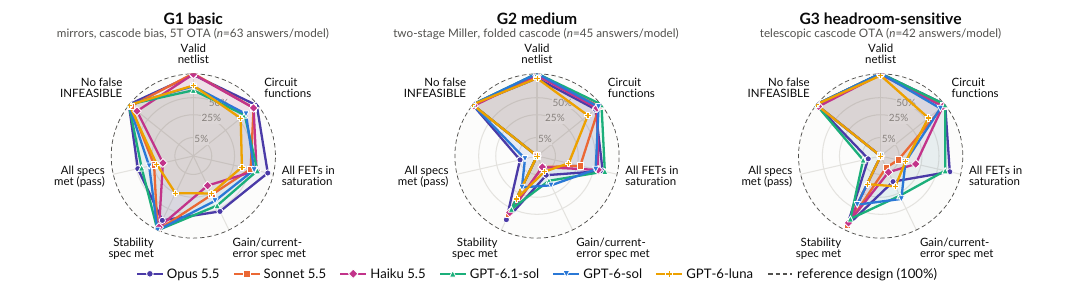}
\smallerspacecaption
\smallerspacecaption
\smallerspacecaption
\smallerspacecaption
\caption{Capability per circuit grade (zero-shot, medium effort, T1--T3); each axis is scored independently on a square-root radial scale.}
\label{fig:grades} \vspace{-10pt}
\end{figure*}

\begin{figure}[t]
\centering
\includegraphics[width=\columnwidth]{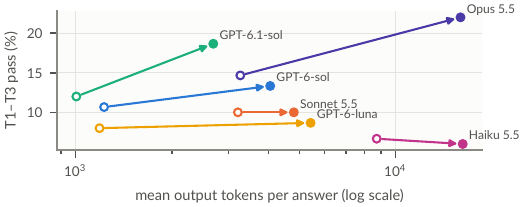}
\smallerspacecaption
\smallerspacecaption
\smallerspacecaption
\smallerspacecaption
\caption{Pooled T1--T3 pass rate versus mean output tokens per answer, from low (open marker) to medium (filled) reasoning effort.}
\label{fig:levers} \vspace{-9pt}
\end{figure}


\textbf{RQ3. Do LLMs recognize infeasible specifications?} This study is conducted on G1 circuits. As shown, Opus~5.5 performs best by flagging 10-11 of the 12 infeasible specifications with almost no false alarms. Haiku~5.5 and GPT-6-luna at low effort flag frequently but also produce false alarms on feasible tasks. 
Increasing reasoning effort did not improve infeasibility detection, which is consistent with findings in~\cite{kirichenko2025abstentionbench}.  Moreover, detection capability is not fully aligned with porting skill, since GPT-6.1-sol, the strongest GPT model, ports well but rarely flags an infeasible specification. It frequently returns designs that silently miss an unreachable limit.

\textbf{RQ4. More reasoning or simulation feedback?} Our studies show that more reasoning helps only the strongest models, while increasing token cost (Fig.~\ref{fig:levers}). 
While Opus~5.5 and GPT-6.1-sol gain sizably, others gain little or nothing despite using up to 4.6$\times$ more tokens. 
In comparison, simulation feedback is useful to all models. 
Since self-repair~\cite{chen2023selfdebug,shinn2023reflexion} may add little over equal-cost resampling~\cite{olausson2023selfrepair}, we compare three feedback attempts against the best of three independent samples. 
For the two strongest models, feedback solves over half of the feasible tasks, against about a third for resampling (Table~\ref{tab:main}). 
Detailed examination indicates that feedback helps most on hard specifications, raising solved T3 tasks from 2 to 13 for GPT-6.1-sol and from 2 to 12 for Opus~5.5.

\textbf{RQ5. Where in the design flow do designs fail?} 
Fig.~\ref{fig:grades} plots design flow steps clockwise. 
On Easy circuits, 
shapes narrow at saturation and current error. 
On Medium circuits, shapes stay wide except on gain, meeting at most 16\% of answers against up to 73\% for stability, suggesting a gain-stability trade-off. 
On Hard circuits, the saturation axis separates models (RQ2), and even designs with all transistors in saturation lose gain.

\vspace{-5pt}
\section{Conclusion}

This work presented \textsc{AnalogPortBench}, a benchmark that evaluates LLMs on porting analog circuits to a new process node and lower supply. 
By evaluating six frontier LLMs, 
we found that models port simple circuits but rarely redesign headroom-sensitive amplifiers, and only one model reliably recognizes infeasible specifications. 
Simulation feedback improves LLM performance more than additional samples or reasoning. 
Open gaps include headroom-aware topology choice, infeasibility recognition, LLM-guided numerical sizing, and corner-robust design.

\clearpage
\newpage
\bibliographystyle{IEEEtran}
\bibliography{refs}

\end{document}